\documentclass[10pt,oneside,a4paper]{article}

\usepackage[utf8]{inputenc}
\usepackage{graphicx}
\usepackage{dcolumn}
\usepackage{bm}
\usepackage{braket}
\usepackage{color}
\usepackage{units}
\usepackage{dsfont}
\usepackage{float}
\usepackage{caption}
\usepackage{hyperref}
\usepackage{amsmath}
\usepackage{xcolor}

\usepackage[noadjust]{cite}
\usepackage{filecontents}

\usepackage[a4paper, total={6.25in, 8.5in}]{geometry}

\DeclareMathOperator{\Trace}{Tr}

\begin{document}

\title{{\Large Near-perfect measuring of full-field transverse-spatial modes of light}}

\author{Markus Hiekkamäki$^{1}$, Shashi Prabhakar$^{1}$ and Robert Fickler$^{1,*}$}
\date{{\footnotesize $^{1}$Photonics Laboratory, Physics Unit, Tampere University, Tampere, FI-33720, Finland \\ $^{*}$robert.fickler@tuni.fi}}

\maketitle



\begin{abstract}
Along with the growing interest in using the transverse-spatial modes of light in quantum and classical optics applications, developing an accurate and efficient measurement method has gained importance. Here, we present a technique relying on a unitary mode conversion for measuring any full-field transverse-spatial mode. Our method only requires three consecutive phase modulations followed by a single mode fiber and is, in principle, error-free and lossless. We experimentally test the technique using a single spatial light modulator and achieve an average error of 4.2~\% for a set of 9 different full-field Laguerre-Gauss and Hermite-Gauss modes with an efficiency of up to 70\%. Moreover, as the method can also be used to measure any complex superposition state, we demonstrate its potential in a quantum cryptography protocol and in high-dimensional quantum state tomography. 
\end{abstract}

\section{Introduction}
Using light to encode and transmit information is key in today’s information technology-driven age. Amongst the different degrees of freedom, encoding information in the transverse-spatial degree of light has attracted significant attention over the last years, as it offers another way to increase bandwidth \cite{Willner}. During these past years, the rates with which information is transmitted using the transverse-spatial structure has been pushed to terabits per second, alongside studies into novel channels to enable long-distance transmissions, such as fibers, free space as well as water \cite{Willner, bozinovic_terabit-scale_2013, krenn2014communication, bouchard_quantum_2018}. 
In addition to their use in classical information technologies, spatial modes have also been harnessed as physical realizations of high-dimensional quantum states \cite{erhard2018twisted}. These so-called \textit{qudits} are beneficial in terms of information capacity per single quantum carrier in addition to their noise resistance in quantum communication schemes \cite{cerf2002security, mirhosseini2015high, ecker2019entanglement}. Moreover, encoding quantum information into the transverse-spatial degree of freedom has also enabled simple quantum simulation and computation schemes \cite{cardano2015quantum, cardano2017detection}.

However, the benefits obtained, when using spatial modes of light in both classical and quantum communication, strongly depend on the ability to generate and measure such modes with high precision and high efficiency. Various techniques have been developed ranging from holographic generation and projection \cite{Heckenberg1992, mair2001entanglement}, to direct transverse phase and amplitude modulation~\cite{beijersbergen1994helical, marrucci2006optical} to mode multi- and de-multiplexing schemes \cite{leach2002measuring, berkhout_efficient_2010, fickler2017custom, ruffato2017test, Fontaine2018:LGsorter, zhou2017sorting, gu2018gouy}, often focusing on the measurement of the modal content of a light-field \cite{schulze2013measurement}. While in some cases sorting or de-multiplexing schemes might be beneficial, it also requires the ability to have multiple detectors or efficient cameras, such that the process of direct filtering or projecting onto modes is necessary. The key idea behind most of such projection techniques relies on the fact that only a Gaussian mode with a plane phase front couple efficiently into single mode fibers (SMF)~\cite{mair2001entanglement}. Thus, when certain light modes are under investigation, the measurement was performed by flattening the transverse phase structures with a spatial light modulator (SLM), which then acts together with the SMF as a mode filter. While this technique works with low modal cross talk for azimuthally structured light fields, it does not work for radially structured fields of light \cite{qassim2014limitations} without inducing a large amount of loss \cite{BouchardPhaseFlat}. Recently, the idea was extended to two phase modulations planes, one in the near and one in the far field, to determine the radial decomposition using a phase-retrieval algorithm \cite{choudhary2018measurement}.

In this article, we demonstrate a technique to project on any type of transverse-spatial mode with, in principle, perfect efficiency and no errors. Our measurement procedure relies on a unitary transformation of any given spatial mode into the Gaussian mode of a single mode fiber implemented via the technique of multi-plane mode conversion. We show that three planes of phase modulation are already enough to detect a broad range of modes, i.e. the mode families of Laguerre-Gauss and Hermite-Gauss modes with errors as low as 2.3~\% and efficiencies reaching values above 70~\%. We further demonstrate the broad applicability of our technique by using it in a 7-dimensional quantum cryptography protocol as well as a full quantum state tomography, where include both the azimuthal as well as the radial transverse structure, i.e. the full-field transverse-spatial modes of light.

\section{Multi-plane mode conversion}
Converting spatial modes of light has been a task, which has not been investigated in a lot of detail over the last years due to its complexity and the non-existence of utilizable bulk optics devices. However, already a few years back it has been shown that multiple phase modulations between free space propagation, can be used to perform some elementary transformations \cite{berkhout_efficient_2010}. 
Although the general working principle was demonstrated, the technique was never extended or implemented in subsequent experiments, possibly due to the complexity of the optimization algorithm used to generate the phase modulations. Only recently, a similar approach using multiple planes of phase modulation was demonstrated that improved the performance by using a technique from waveguide design called wave-front matching (WFM) to obtain the required transverse phase modulations \cite{Hashimoto2005}. This technique compares the complex transverse amplitudes of the input and output light fields and iteratively matches the wave front by simple phase modulations. While it was shown that this technique works for multiple input modes and spatially separated output channels \cite{Fontaine2018:LGsorter} as well as multiple output modes \cite{Brandt2019}, we restrict the iterative process to match one input mode, i.e. the mode to be measured, and a Gaussian output mode, i.e. the mode of the single mode fiber.

The general iterative optimization process is the following: 
The mode to be measured, i.e. the input mode $M$, is propagated forward through an optical system containing $n$ phase elements $\Phi_t$, each followed by some free space propagation. At these modulation planes, $t=1,...,n$, the complex amplitudes of the mode $M(x,y,t)$ is recorded. Then, the Gaussian output mode $G$, i.e. the collimated beam from the single mode fiber, is propagated backwards through the system to the last phase modulation plane ($t=n$) to obtain $G(x,y,n)$. The two fields, $M(x,y,t)$ and $G(x,y,t)$ are now matched by imprinting a transverse phase. This phase is calculated by the field overlap between the mode pair 
\begin{equation}
o_{t}(x,y)=\overline{M(x,y,t)} G(x,y,t) e^{i\Phi_t(x,y)},    
\end{equation}
including a transverse phase modulation $\Phi_t(x,y)$, which is set to zero in the beginning, but will be updated during the WFM process. The transverse phase of this overlap, offset by its mean value $\phi$, is then the required change in the phase pattern for the plane $t$, i.e. 
\begin{equation}
\label{eq:phaseupdate}
\Delta\Phi_t(x,y) = - \arg (o_{t}(x,y)e^{-i\phi}).
\end{equation} 
This phase modulation is imprinted on the backwards propagating Gaussian mode $G$ and subsequently propagated to the $(n-1)$th modulation plane. We note that due to the free-space propagation between two modulation planes, the amplitude is also slowly adjusted to match the mode $M$ to the Gaussian output mode. At plane $(n-1)$, the same procedure is repeated, i.e. both modes are “compared” and matched by another phase modulation, before the field $G$ is further propagated backwards. This procedure is repeated till the first plane. If the number of phase modulation planes is large enough, one such optimization already matches the input mode perfectly to the Gaussian mode, i.e. it acts in forward direction as a unitary mode transformation. If now a single mode fiber, which only allows coupling of Gaussian modes, is placed into the beam after this transformation, a perfect projection onto the mode $M$ is performed. This exclusive coupling naturally arises due to the unitarity of the transformation, which preserves the orthogonality of the modes.

If the number of phase modulations is limited, the whole procedure can be repeated, also in the forward propagation, until a desired fidelity is achieved. In this regard, in equation \ref{eq:phaseupdate}, $\phi$ is used to add an offset to each phase modulation plane, which shortens the convergence time for the optimization using multiple iterations. While generating the phase modulations we found that, for all the modes we investigated, three phase modulations were enough to achieve an overlap of 99.9~\% between the converted mode and a Gaussian. An example of the phase transformation obtained for transforming a LG mode with indices $l=-1, p=1$ into a Gaussian mode can be seen in Fig.\ref{fig1}.

\section{Experimental setup}
After having established a measurement method that relies on a unitary transformation between any given input mode and a Gaussian mode plus a coupling into a single mode fiber, we now test how current technical limitations affect the quality of the measurement scheme.

\begin{figure}[ht]
    \centering
    \captionsetup{width=0.8\textwidth}
    \hspace{-0.5cm}
    \includegraphics[width = 13cm]{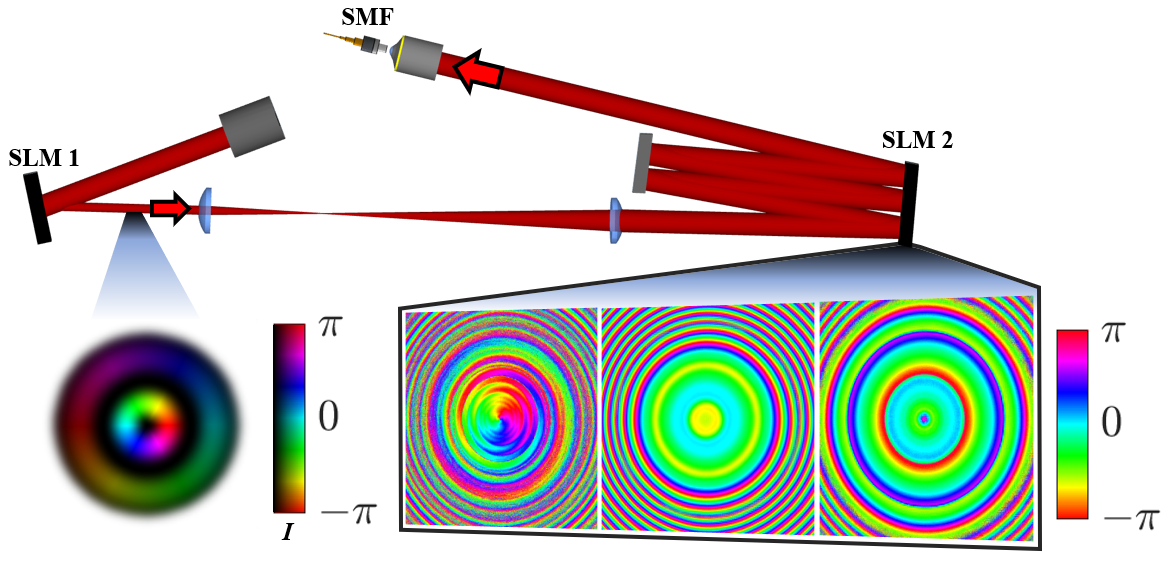}
    \caption{Sketch of the experimental setup for near-perfect spatial mode measuring. A spatial light modulator (SLM1) was used to generate a specific mode to be measured. A magnifying imaging system enlarges the beam size and thus the number of pixels illuminated in the mode transformation is increased. Three phase modulations on SLM2 transform the input mode into a Gaussian, which exclusively couples into the single mode fiber (SMF). The insets display a LG mode $l=-1, p=1$ (lower left) and the required transforming phase modulations (lower right) as an example. Note that for the mode generation we also use an amplitude masking scheme \cite{Bolduc:ModeCarving} and overlay all phase structures with additional gratings in the experiment  (not shown).}
    \label{fig1}
\end{figure}

We use a simple experimental setup (see Fig. \ref{fig1}), where we at first generate any complex transverse light field using SLM1 and shape the incident laser beam (808 nm) by modulating its transverse phase and amplitude through a complex and lossy hologram \cite{Bolduc:ModeCarving}. We then redirect the modulated beam into our mode measurement system, consisting of mode-converter and a single mode fiber. We realize the mode-converter using multi-plane phase modulations displayed on SLM2 on three separate regions. To achieve this, we place a mirror 40 cm away from SLM2 parallel to its surface. For experimental convenience, we also adjust the beam waist during the mode conversion procedure to match it to the beam waist we measured for our coupling system, i.e. a microscope objective (20x) and a SMF. To reduce misalignment errors and detrimental effects due to the finite resolution of our SLM in the mode conversion (Holoeye, 8$\mu$m pixel pitch), we use a beam waist of 0.94 mm at the input of our measurement system and phase modulations spanning 630 by 630 pixels. As the phase modulation using an SLM is only 75~\% efficient, we additionally display a blazed grating structure in all of the phase modulations and only use the first diffraction order from each phase screens. Note, that this additional diffraction is only required due to the limited efficiency of an SLM, which can be overcome by using custom-designed diffractive optical phase elements \cite{ruffato2017test}.

\section{Measurement of full-field transverse-spatial modes}
We first test the presented method by measuring the nine lowest order modes of two of the most common mode families, i.e. the Laguerre-Gauss (LG) and Hermite-Gauss (HG) modes. 

\subsection{Laguerre-Gauss modes}
The LG mode family is obtained by solving the paraxial wave equation in cylindrical coordinates \cite{andrews_angular_2013}. The modes form a complete orthonormal set, where the first index $l$ corresponds to the azimuthal structure and the second index $p$ describes the radial profile. These modes have attracted a lot of attention over the last decades as they are nicely matching the symmetry of most of the optical devices and, more importantly, the azimuthal index $l$ corresponds to an orbital angular momentum (OAM) caused by the twisted phase front $e^{i l \varphi}$, where $\varphi$ is the angular position \cite{allen_orbital_1992, padgett2017orbital}. They have found a myriad of applications, e.g. in optical tweezers\cite{padgett2011tweezers}, optical communication \cite{Willner} as well as quantum optics \cite{erhard2018twisted}. LG modes, and in particular the OAM quantum number ($l$), have been used in various fundamental studies \cite{fickler2016quantum, erhard2018experimental} as well as in quantum information applications \cite{mirhosseini2015high, zhang_experimentally_2016, bouchard_quantum_2018}, where they are used as a physical realization of high-dimensional Hilbert spaces. In our experiment, we test the nine lowest order modes including both the azimuthal and radial indices. 

\begin{figure}[ht]
    \centering
    \hspace{-0mm}
    \captionsetup{width=0.8\textwidth}
    \includegraphics[width = 130mm]{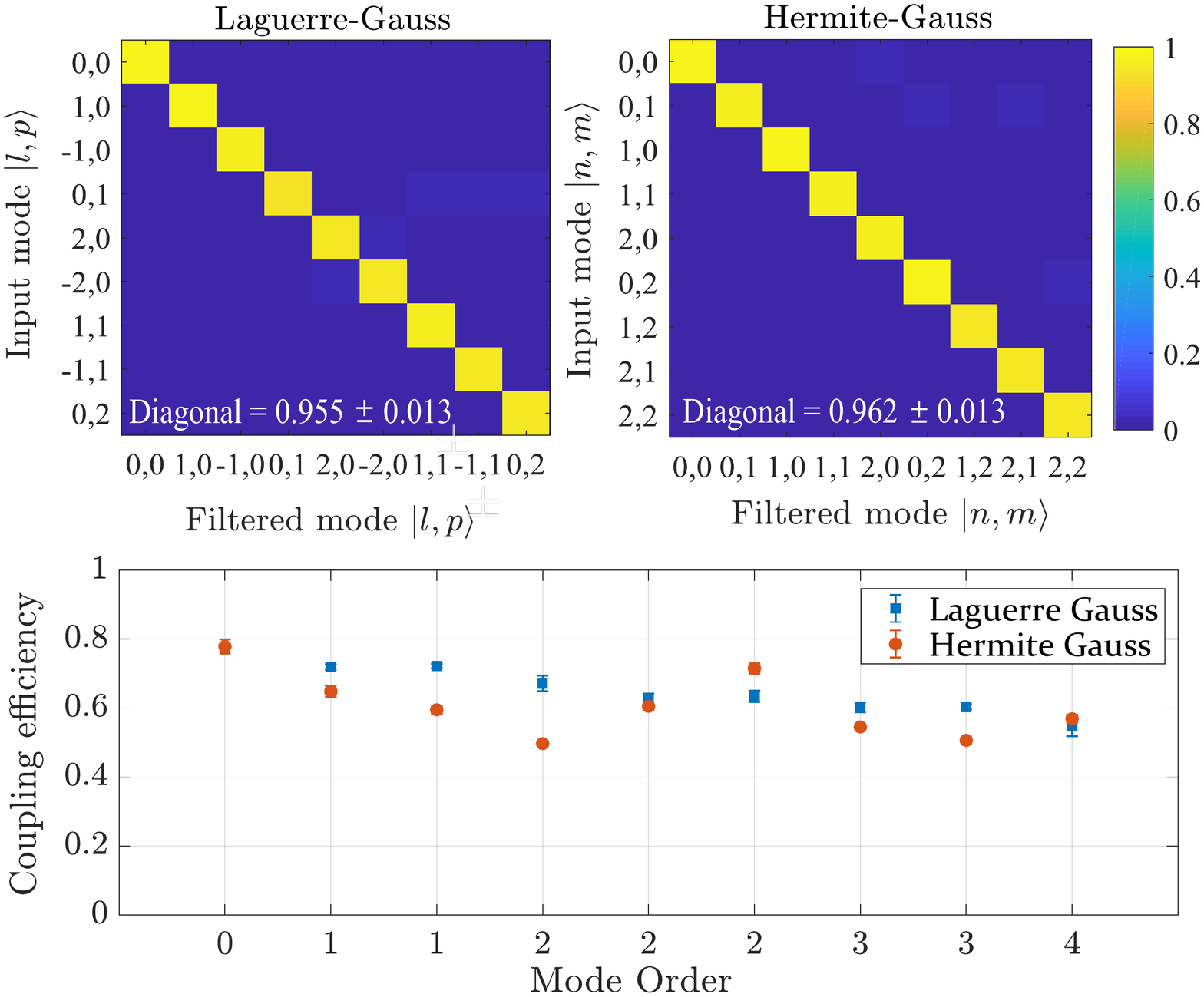}
    \caption{Modal cross talk matrices for the projection measurements of the 9 lowest order a) Laguerre-Gauss and b) Hermite-Gauss modes. In graph c), the SMF coupling efficiencies of the converted modes are shown with the same x-axis as used in a) and b). The given diagonal values show the mean and standard deviation of the normalized detection rates on the cross talk matrix diagonal.}
    \label{fig2}
\end{figure}

Irrespective of the mode order (besides the zeroth order, i.e. Gaussian mode), we find that all modes couple into the SMF after being transformed into a Gaussian mode with an efficiency between 55-72~\%. We note that due to the limited modulation efficiency of the SLM and the three consecutive phase modulations, around 25~\% of the input light was detected after the fiber. A more important measure for many applications is the modal cross talk between all modes, which can be characterized by the visibility $V=\sum_{i}C_{ii}/\sum_{ij}C_{ij}$, where $C_{ij}$ corresponds to the cross-talk matrix (see Fig. \ref{fig2} a). We achieve a visibility of $V_{LG}=95.5 \pm 0.9$~\% for the nine lowest order LG modes. We attribute the reduced efficiency (simulations predict 99.9~\%) and small cross-talk to the finite resolution and some minor miss-alignments. In addition, we generate the modes through amplitude and phase modulation \cite{Bolduc:ModeCarving}, a very good but also not perfect technique. As our multi-plane mode conversion measurement is calculated for perfect modes, the imperfections from generation might also induce errors and cause the slightly reduced coupling. Nevertheless, both results, the high coupling efficiency and visibility, show that the transformation is very close to a unitary operation and, thus, a near-perfect projection.

\subsection{Hermite-Gauss modes}
Another popular mode family is obtained by solving the paraxial wave equation in Cartesian coordinates, i.e. the HG modes. Similarly to the LG modes, HG modes are characterized by two mode indices commonly labeled as $n$ and $m$, which correspond to the number of vertical and horizontal $\pi$ phase jumps, respectively. As our measurement technique works independent of the input mode, we find very similar results, when generating and measuring the lowest, nine order HG modes. Here, the visibility is found to be $V_{HG}=96.2 \pm 1.0$~\% and the efficiency is again around 50-72~\%, independent of the mode order (see Fig. \ref{fig2} c). This first set of measurements shows that, compared to other techniques, projecting on specific modes through the presented technique offers a highly efficient and mode independent way to measure the full-field modal content of a light field with low errors.

\section{Applications in high-dimensional quantum information}
In the second set of experiments, we take advantage of the introduced projection method and studied two important applications in quantum information, namely performing a high-dimensional quantum cryptography protocol and quantum state tomography. Although we could have performed both tasks in the LG or HG mode basis (or any other orthogonal mode set), we perform both tasks using the set of 7 LG modes 
\begin{equation}
\label{eq:7Modes}
\ket{\psi_n}\in\{\ket{-1,1},\ket{-1,0},\ket{0,0},\ket{0,1},\ket{0,2},\ket{1,0},\ket{1,1}\} \equiv \{ \ket{1}, \ket{2}, ..., \ket{7}\},
\end{equation}
where the positions in the ket-vectors label the $l$ and $p$ indices of the LG modes, respectively. We chose the LG mode family as they have been the key player in high-dimensional quantum optics using spatial modes as $d$-dimensional encoding.

\subsection{Quantum Cryptography}
\label{sec:QCrypt}
To test the applicability of our method, we measure the performance of our mode filter if applied to a quantum cryptography scheme, in particular the high-dimensional version of the well-known BB84 protocol \cite{bouchard_experimental_2018}. This protocol requires measurements in two mutually unbiased bases, between which, the two parties randomly select states to establish a secure key. In our test, one basis will be realized by the set of seven LG modes defined in equation \ref{eq:7Modes}. The second basis set is in a mutually unbiased basis (MUB), and is obtained through the linear superpositions $\ket{\varphi_n} = \frac{1}{\sqrt{d}} \sum_{m=0}^{d-1}\omega^{nm}_d\ket{\psi_m}$, with $\omega_d=\exp{(i2\pi/d)}$, and is often also called Fourier-basis as it is obtained through the quantum Fourier transformation. 

\begin{figure}[ht]
    \centering
    \captionsetup{width=0.8\textwidth}
    \hspace{-0cm}
    \includegraphics[width = 9.5cm]{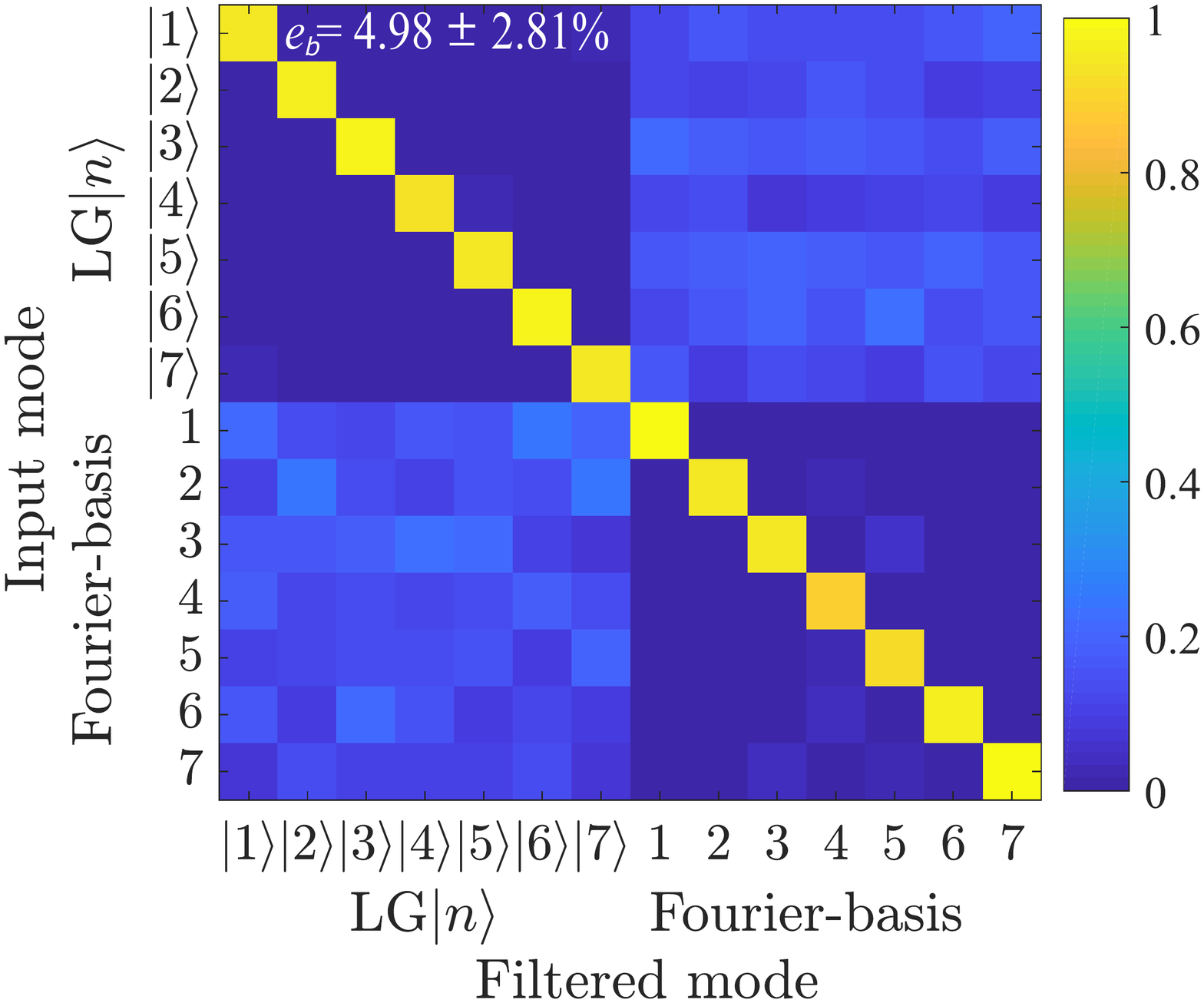}
    \caption{Modal cross talk matrix of the 14 modes used in a 7-dimensional quantum cryptography protocol. The secret key rate calculated from the bit error rate is 1.98 bits per sifted photon.}
    \label{fig3}
\end{figure}

Using qudits instead of bi-dimensional systems in quantum cryptography allows one to transmit more information per carrier, i.e. $\log_2(d)$-bit per photon for the error-free case. Additionally, such protocols tolerate larger error rates, such that they are especially useful in noisy conditions. The secret key rate $R$ is given by $R = \log_2(d) - 2h^{(d)}(e_b)$, where $e_b$ is the bit error rate and $h^{(d)}(x):=-x\log_2(x/(d-1))-(1-x)\log_2(1-x)$ is the $d$-dimensional Shannon entropy.

The obtained cross talk matrix can be found in Fig. \ref{fig3}. The average error rate was found to be 4.98$\pm$ 2.81~\%, which corresponds to $R=$1.98 bits per sifted photon. We note that by using a single outcome measurement like ours instead of a multi-outcome measurement scheme, the overall efficiency, i.e. the bit rate per sent photon, is reduced by a factor of $d$. Nevertheless, our measurement technique allows for a bit rate that is larger than 1 bit per photon using both, the azimuthal $l$ and radial $p$ degree of freedom of the photons.

\subsection{Quantum State Tomography}
\begin{figure}[t]
    \centering
    \captionsetup{width=0.8\textwidth}
    \hspace{-0cm}
    \includegraphics[width = 13cm]{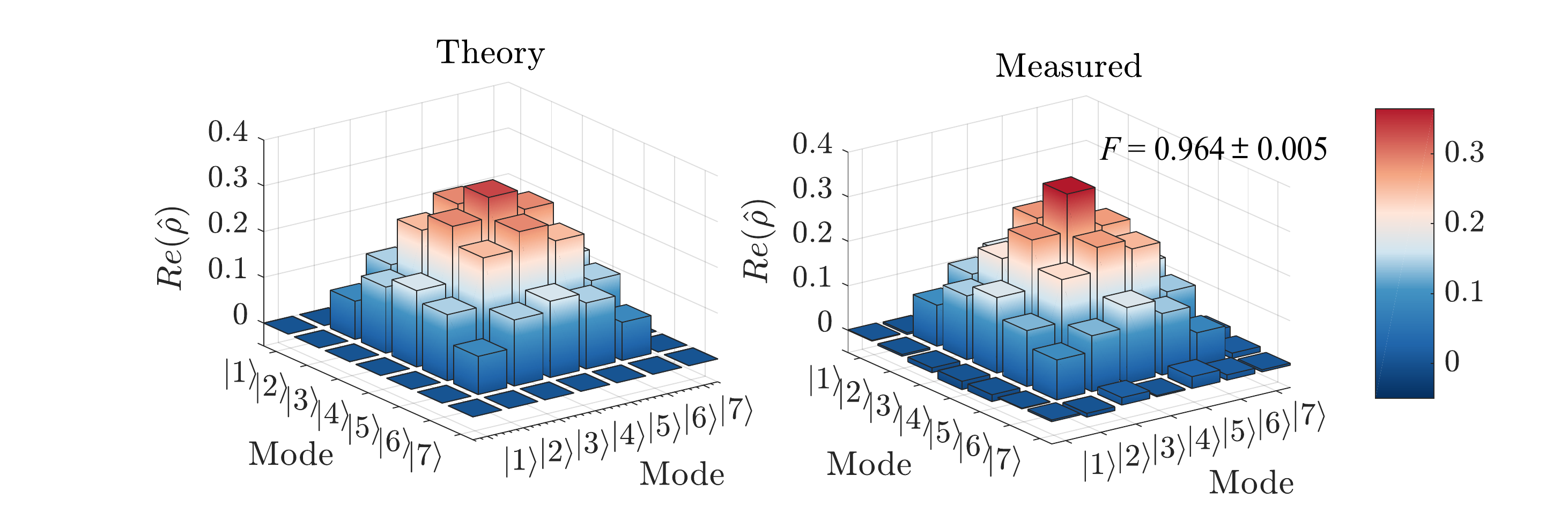}
    \caption{Real part of the density matrix reconstructed using quantum state tomography.}
    \label{fig4}
\end{figure}

Another important task in quantum information schemes is the precise measurement of a quantum state. Through the so-called quantum state tomography, it is possible to reconstruct the full density matrix $\hat{\rho}=\ket{\Psi}\bra{\Psi}$ and, thus, fully characterize a generated state. One way to perform such state tomography is to measure the state in all MUBs. In prime and power of prime dimensions, the number of MUBs is known to be $d$+1 \cite{durt2010mutually}. We again use the same 7-dimensional LG mode state space described in section \ref{sec:QCrypt} and generate a visually appealing state of the form $\ket{\Psi} = \frac{1}{N} \sum_{n=0}^{6} \sin(\frac{n\pi}{6}) \ket{n+1}$, where $N$ is a normalization constant and $\ket{n}$, are the LG modes as defined in equation \ref{eq:7Modes}. In addition to the computational basis states given in equation \ref{eq:7Modes}, the $k$ MUBs to perform the high-dimensional state tomography can be constructed through the high-dimensional Hadamard transformations
\begin{eqnarray}
\ket{\phi_n^{(k)}} = \frac{1}{\sqrt{d}} \sum_{m=0}^{d-1}\omega^{(nm+(k-1)m^2)}_d\ket{\psi_m},
\label{eq:Hadamard}
\end{eqnarray}
again with $\omega_d=\exp{(i2\pi/d)}$. For $k$=1, equation (\ref{eq:Hadamard}) leads to the above-described quantum Fourier transform, i.e. Fourier-basis. After performing all $d(d+1)$ measurements, to avoid systematic errors \cite{schwemmer2015systematic}, we reconstruct the density matrix through direct inversion given by $\hat{\rho}=\sum_{k,n}P_n^{(k)}\Pi_m^{(k)}-\mathds{1}$, where $P_n^{(k)}$ corresponds to the probability of measuring the $n$-th state of the $k$-th MUB, i.e. $\ket{\phi_n^{(k)}}$, and $\Pi_m^{(k)}$ corresponds to the projector onto that state, i.e. $\ket{\phi_n^{(k)}}\bra{\phi_n^{(k)}}$. The results of this reconstruction can be seen in Fig. \ref{fig4}. 

As a good measure to evaluate how close the measured density matrix $\hat{\rho}_{exp}$ is with respect to the theoretical density matrix $\hat{\rho}_{th}$, we use fidelity defined as 
\begin{equation}
F=\left(\Trace \sqrt{\sqrt{\hat{\rho}_{exp}}\hat{\rho}_{th}\sqrt{\hat{\rho}_{exp}}}\right)^2.
\end{equation}
From our measurement we achieve a fidelity of 96.4 $\pm$ 0.5~\% for our reconstructed state, which again shows the quality of the projection method introduced here.

\section{Conclusion and Outlook}
In conclusion, we have presented a measurement technique that relies on mode conversion requiring only three phase modulating planes to perform a near-perfect unitary transformation of any given transverse-spatial mode into a Gaussian mode and a subsequent coupling into a single mode fiber. We achieved very high efficiencies irrespective of the mode order as well as very low cross talk between other modes. We further have demonstrated the quality of the projections for two very prominent mode families, i.e. the Laguerre-Gauss and Hermite-Gauss modes, and applied the techniques in a high-dimensional quantum cryptography protocol and also performed a quantum state tomography. 

As our technique is highly efficient and can measure the full field structure, it can be directly applied in quantum optics experiments, e.g. to enable the measurements of correlations in the azimuthal and radial degree of freedom of a high-dimensional entangled bi-photon state generated in down-conversion experiments \cite{mair2001entanglement, ecker2019entanglement}. Moreover, it can be used to decompose the transverse light field into any specific mode basis of choice and as such might be applied in a broad range of experiments investigating the spatial domain. Finally, as the transformation is unitary it can also be used in reverse to perfectly generate complex spatial modes or a desired transverse structure of light in a highly efficient, near-perfect manner.

\section*{Funding Information}
MH, SP and RF acknowledge the support of the Academy of Finland through the Competitive Funding to Strengthen University Research Profiles (decision 301820) and the Photonics Research and Innovation Flagship (PREIN - decision 320165).

\bibliography{ref}
\bibliographystyle{IEEEtran}

\end{document}